\documentclass[sigplan]{acmart}

\renewcommand\footnotetextcopyrightpermission[1]{}
\usepackage{booktabs}
\usepackage{multirow}
\usepackage{pifont}
\usepackage{listings}
\usepackage{xcolor}
\usepackage[capitalise,noabbrev]{cleveref}
\usepackage[ruled,linesnumbered,noend]{algorithm2e}
\crefname{algocf}{Algorithm}{Algorithms}
\Crefname{algocf}{Algorithm}{Algorithms}
\usepackage[most]{tcolorbox}
\usepackage{pgfplots}
\pgfplotsset{compat=1.18}
\usepgfplotslibrary{statistics}
\usetikzlibrary{positioning}

\newtcolorbox{findingbox}[1][]{%
  enhanced,
  colback=black!4,
  colframe=black!50,
  boxrule=0.6pt,
  left=6pt, right=6pt, top=4pt, bottom=4pt,
  fontupper=\small\itshape,
  before skip=4pt,
  after skip=6pt,
  #1
}

\definecolor{jsonkey}{rgb}{0.0, 0.0, 0.6}
\definecolor{jsonstring}{rgb}{0.0, 0.5, 0.0}
\lstdefinelanguage{json}{
  basicstyle=\ttfamily\small,
  string=[s]{"}{"},
  stringstyle=\color{jsonstring},
  literate=
    *{:}{{\color{jsonkey}:}}{1}
    {,}{{\color{black},}}{1}
    {\{}{{\color{black}\{}}{1}
    {\}}{{\color{black}\}}}{1}
    {[}{{\color{black}[}}{1}
    {]}{{\color{black}]}}{1},
  breaklines=true,
  frame=single,
  framerule=0.5pt,
}

\lstdefinelanguage{cpgql}{
  basicstyle=\ttfamily\small,
  keywords={cpg, method, name, parameter, call, reachableByFlows, filter, where, local, assignment, code, lineNumber, l, toJson},
  keywordstyle=\color{blue!70!black},
  stringstyle=\color{jsonstring},
  commentstyle=\color{gray},
  morestring=[b]",
  breaklines=true,
  frame=single,
  framerule=0.5pt,
}

\newcommand{\approach}[1]{\textsc{A#1}}
\newcommand{\cpgql}{\textsc{CPGQL}}

\begin{document}

\title{Less Is More: Measuring How LLM Involvement\\Affects Chatbot Accuracy in Static Analysis}

\author{Krishna Narasimhan}
\email{krishna.nm86@gmail.com}
\affiliation{%
  \institution{F1Re BV}
  \city{Amsterdam}
  \country{The Netherlands}
}
\begin{abstract}
Large language models are increasingly used to make static analysis tools
accessible through natural language, yet existing systems differ in how
much they delegate to the LLM without treating the degree of delegation
as an independent variable. We compare three architectures along a
spectrum of LLM involvement for translating natural language to Joern's
query language \cpgql{}: direct query generation (\approach{1}),
generation of a schema-constrained JSON intermediate representation
(\approach{2}), and tool-augmented agentic generation (\approach{3}).
These are evaluated on a benchmark of 20 code analysis tasks across
three complexity tiers, using four open-weight models in a 2\(\times\)2
design (two model families \(\times\) two scales), each with three
repetitions. The structured intermediate representation (\approach{2})
achieves the highest result match rates, outperforming direct generation
by 15--25 percentage points on large models and surpassing the agentic
approach despite the latter consuming 8\(\times\) more tokens.
The benefit of structured intermediates is most pronounced for large
models; for small models, schema compliance becomes the bottleneck.
These findings suggest that in formally structured domains, constraining
the LLM's output to a well-typed intermediate representation and
delegating query construction to deterministic code yields better results
than either unconstrained generation or iterative tool use.
\end{abstract}



\maketitle

\section{Introduction}\label{sec:intro}

Knowledge about software systems exists in two fundamentally different forms.
On one side is the unstructured world: natural language requirements, developer questions posed in issue trackers, security policies described in prose, vulnerability reports written for human readers.
On the other side is the structured world: formal query languages, typed schemas, deterministic program representations, machine-executable specifications.
Bridging these two forms is a recurring problem in software engineering, and the difficulty of the crossing depends on how far apart the two sides are.
Text-to-SQL translation~\cite{spider2018,bird2023} bridges natural language questions and relational query languages; natural-language-to-code generation~\cite{chen2021codex} bridges task descriptions and executable programs; requirements formalisation bridges prose specifications and formal models.
In each case, the core challenge is the same: an intent expressed informally must be rendered in a language with precise syntax and semantics.

Static analysis tools sit firmly on the structured side.
Tools like Joern~\cite{joern2014} construct Code Property Graphs (CPGs) that unify abstract syntax trees, control flow graphs, and data flow graphs into a single queryable structure, and expose \cpgql{}, a Scala-flavoured traversal language for querying this representation.
\cpgql{} is powerful---it supports structural pattern matching, inter-procedural data flow tracing, and call graph resolution---but writing queries requires fluency in a niche DSL with method chaining, graph traversal semantics, and knowledge of the CPG schema.
Despite well-documented benefits of static analysis, adoption remains limited by usability rather than capability: developers struggle with warning messages~\cite{nachtigall2022}, false positives discourage routine use~\cite{johnson2013}, and integrating these tools into workflows requires repeated design iterations~\cite{sadowski2018}.
The query interface is a key source of this friction.

Large language models have emerged as candidate bridges between the unstructured and structured worlds~\cite{hou2024llmse}.
An LLM can interpret a natural language request and produce output that approximates a structured artefact---a database query, a code snippet, a configuration file~\cite{chen2021codex}.
But LLMs are stochastic approximations, not faithful translators.
They pattern-match against distributional regularities in training corpora rather than reason over formal semantics, and they hallucinate plausible-looking outputs that may be syntactically valid but semantically wrong~\cite{liu2024evalplus}.
Their reliability degrades further on formalisms that are underrepresented in training data~\cite{structeval2025,dslsurvey2025}, which includes most domain-specific languages encountered in practice.
The question, then, is not whether to use LLMs for this translation, but how much latitude to give them: how tightly to constrain their output space while still leveraging their capacity for language understanding.

Several systems explore LLM-based translation to address the static analysis usability gap.
IRIS~\cite{iris2025} restricts the LLM to labelling APIs as taint sources or sinks, keeping query generation entirely deterministic.
QLCoder~\cite{qlcoder2025} uses an agentic framework to synthesise full CodeQL queries but reports that giving the agent unrestricted compile-and-run access degraded performance.
MoCQ~\cite{mocq2025} reduces the DSL surface exposed to the model, finding that the full specification overwhelmed it.
An earlier position paper proposed a translation-based architecture for Joern specifically~\cite{narasimhan2024benevol}.
These systems differ in how much they delegate to the LLM, but none treats the degree of delegation as a variable worth studying.

There is empirical reason to be cautious about delegation.
Anand et al.~\cite{anand2024critical} analysed attention maps and hidden representations in code-LLMs and found that these models encode relations among syntactic tokens and among identifiers, but fail to encode the relations \emph{between} them.
This cross-category reasoning---connecting a keyword like \texttt{if} to the variable it tests, or linking a function call to the data that flows into it---is precisely what static analysis queries require.
The implication is that LLMs lack the internal representations needed to generate reliable analysis queries, even when their output looks syntactically plausible.

This paper treats LLM involvement as an independent variable.
Three architectures are compared along a spectrum of LLM autonomy:
\begin{description}
  \item[\approach{1}: Direct Generation.] The LLM generates \cpgql{} queries directly, aided by retrieval-augmented context from Joern documentation and examples.
  \item[\approach{2}: Structured Intermediate.] The LLM produces a JSON object conforming to a fixed schema. A deterministic mapper translates this to the correct \cpgql{} invocation. The LLM never sees query syntax.
  \item[\approach{3}: Tool-Augmented Agentic.] Analysis capabilities are exposed as tools via function calling. The LLM selects tools and provides arguments in a multi-step loop, following the ReAct pattern~\cite{yao2023react}.
\end{description}

These architectures are evaluated on a purpose-built benchmark of 20 code analysis tasks across three complexity tiers (structural, data flow, composite), using four open-weight models in a 2\(\times\)2 design---two model families (Llama, Qwen) at two scales (7--8B, 70--72B)---with three repetitions each, yielding 720 trials.
The core finding is that the structured intermediate representation (\approach{2}) achieves the highest result correctness, outperforming both direct generation and tool-augmented agentic generation.
The benefit is most pronounced for large models (15--25 percentage points (pp) improvement over \approach{1}), while for small models the gain is modest (3--5\,pp) and limited by schema compliance failures.
The agentic approach (\approach{3}) performs worst despite consuming approximately 8\(\times\) more tokens per task.

The contributions of this paper are:
\begin{enumerate}
  \item A benchmark of 20 natural-language-to-\cpgql{} translation tasks across three complexity tiers, machine-validated against Joern 4.0 and publicly available.
  \item A controlled comparison of three LLM architectures representing different points on the autonomy spectrum, evaluated with uniform retry policies for fair comparison.
  \item Empirical evidence that constraining LLM output to a structured intermediate representation outperforms both direct generation and tool-augmented agentic approaches for DSL generation in a formally structured domain.
  \item Evidence of a model-size interaction: the benefit of structured intermediates is strongest for large models, while small models face a different bottleneck in schema compliance.
\end{enumerate}

\section{Approach}\label{sec:approach}

This section first describes LLM involvement as a design variable and why it matters (\Cref{sec:involvement}), then details the three architectures (\Cref{sec:architectures}).

\subsection{LLM Involvement as a Design Variable}\label{sec:involvement}

Consider a system that takes a natural language request (e.g., ``trace how user input reaches the database call in processOrder'') and produces the output of a code property graph query---a list of methods, a set of data flow paths, a collection of matching code locations.
The system has two parts: an LLM that interprets the request, and deterministic code that executes the query.
The LLM produces some intermediate output, and the deterministic code consumes it.

The design question is: what should that intermediate output look like?

In \approach{1}, the intermediate output is a \cpgql{} query string.
The LLM must produce syntactically and semantically valid Scala-flavoured DSL code, a language with minimal representation in training corpora.
The output space is effectively unbounded: any string could be a query attempt.

In \approach{2}, the intermediate output is a JSON object conforming to a fixed schema with five query types, ten possible output columns, and three flow endpoint types.
The output space is compact and well-typed.
A deterministic mapper---verified by unit tests---translates valid JSON into the correct \cpgql{} invocation.

In \approach{3}, the intermediate output is a sequence of tool invocations.
At each step, the LLM selects a tool and provides arguments.
The output space per step is bounded (a finite set of tools with typed arguments), but errors compound across steps.
A wrong tool choice at step $i$ can derail all subsequent steps.

The pattern is: the smaller and more constrained the intermediate output space, the less can go wrong.
This holds as long as the deterministic mapping from constrained output to domain action is implemented correctly, which is a reasonable assumption when the domain provides schemas and validation.

This argument does not apply universally.
In exploratory domains where the mapping from intent to action is not known upfront, the LLM's ability to plan and improvise has value.
The claim here is limited to domains with sufficient formal structure, which includes static analysis query languages.

\subsection{Direct, Structured, and Agentic Generation}\label{sec:architectures}

\begin{figure*}[t]
\centering
\begin{tikzpicture}[
    node distance=8mm,
    box/.style={draw, rounded corners=2pt, font=\small,
                minimum width=32mm, minimum height=9mm,
                align=center, inner sep=3pt},
    llm/.style={box, fill=black!15},
    det/.style={box, fill=white},
    io/.style={font=\small, align=center, inner sep=1pt},
    arr/.style={->, >=stealth, semithick},
    lbl/.style={font=\small\bfseries, anchor=south},
    dasharr/.style={->, >=stealth, semithick, dashed},
]
\begin{scope}[xshift=0cm]
  \node[lbl] (l1) at (0,0) {\approach{1}: Direct};
  \node[io, below=3mm of l1] (a1in) {NL request};
  \node[llm, below=4mm of a1in] (a1llm) {LLM generates\\[-1pt]CPGQL string};
  \node[det, below of=a1llm] (a1run) {Joern executes};
  \node[io, below=4mm of a1run] (a1out) {Results};
  \draw[arr] (a1in) -- (a1llm);
  \draw[arr] (a1llm) -- (a1run);
  \draw[arr] (a1run) -- (a1out);
  \draw[dasharr] (a1run.east) -- ++(5mm,0) |- node[right, font=\small, pos=0.25]{retry} (a1llm.east);
\end{scope}
\begin{scope}[xshift=5.5cm]
  \node[lbl] (l2) at (0,0) {\approach{2}: Structured};
  \node[io, below=3mm of l2] (a2in) {NL request};
  \node[llm, below=4mm of a2in] (a2llm) {LLM generates\\[-1pt]JSON $\in$ schema};
  \node[det, below of=a2llm] (a2map) {Mapper $\to$ CPGQL};
  \node[det, below of=a2map] (a2run) {Joern executes};
  \node[io, below=4mm of a2run] (a2out) {Results};
  \draw[arr] (a2in) -- (a2llm);
  \draw[arr] (a2llm) -- (a2map);
  \draw[arr] (a2map) -- (a2run);
  \draw[arr] (a2run) -- (a2out);
\end{scope}
\begin{scope}[xshift=11cm]
  \node[lbl] (l3) at (0,0) {\approach{3}: Agentic};
  \node[io, below=3mm of l3] (a3in) {NL request};
  \node[llm, below=4mm of a3in] (a3sel) {LLM selects\\[-1pt]tool + args};
  \node[det, below of=a3sel] (a3exe) {Tool executes};
  \node[llm, below of=a3exe] (a3int) {LLM interprets\\[-1pt]result};
  \node[io, below=4mm of a3int] (a3out) {Answer};
  \draw[arr] (a3in) -- (a3sel);
  \draw[arr] (a3sel) -- (a3exe);
  \draw[arr] (a3exe) -- (a3int);
  \draw[arr] (a3int) -- (a3out);
  \draw[dasharr] (a3int.east) -- ++(5mm,0) |- node[right, font=\small, pos=0.25]{loop} (a3sel.east);
\end{scope}
\end{tikzpicture}
\caption{The three architectures. Grey boxes are LLM-mediated; white boxes are deterministic.}
\label{fig:overview}
\end{figure*}
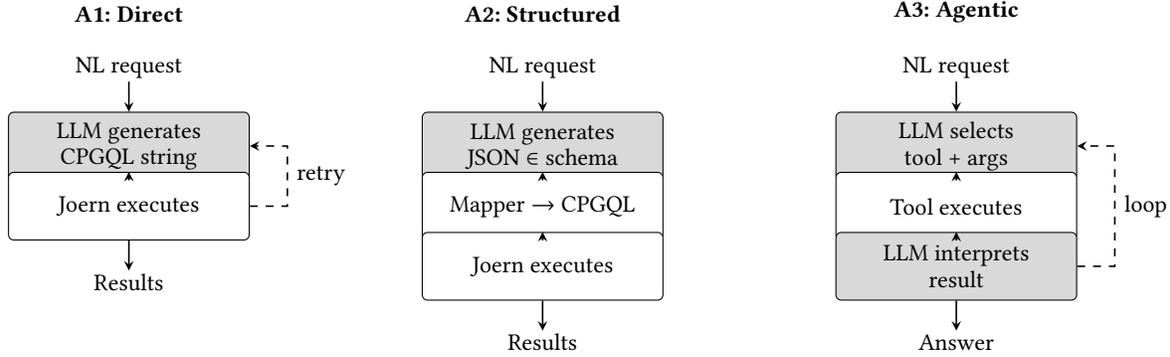

\Cref{fig:overview} shows the three architectures side by side.
In each, the user provides a natural language description of a code analysis task, and the system produces results from Joern's code property graph.
All three share a uniform retry policy: up to three attempts on recoverable errors, with error messages fed back to the LLM.
This ensures a fair comparison that isolates the effect of the architectural choice from the error-correction budget.
\Cref{fig:algorithms} summarises the procedures.

\subsubsection{Approach 1: Direct Generation}\label{sec:a1}

The LLM receives the user request together with a \cpgql{} syntax reference, Joern-specific conventions (operator naming patterns, annotation matchers, output formatting idioms), and worked examples drawn from Joern documentation.
It produces a \cpgql{} string, which is sent to the Joern server for execution.
On execution failure, the error is appended to the conversation history and the LLM is asked to correct its query.

\begin{figure*}[t]
\centering
\begin{minipage}[t]{0.31\textwidth}
\centering
\footnotesize
\textbf{Algorithm 1} \approach{1}: Direct generation.\label{alg:a1}\par\smallskip
\begin{tabular}{@{}l@{}}
\textbf{Input:} NL task $t$ \quad \textbf{Output:} Result or failure\\[2pt]
$\mathit{msgs} \gets [\text{sys}(\text{ref}), \text{usr}(t)]$\\
\textbf{for} $i \gets 1$ \textbf{to} $3$ \textbf{do}\\
\quad $q \gets \mathrm{extract}(\mathrm{LLM}(\mathit{msgs}))$\\
\quad $(\mathit{ok}, r, e) \gets \mathrm{Joern}(q)$\\
\quad \textbf{if} $\mathit{ok}$ \textbf{then return} $r$\\
\quad $\mathit{msgs} \mathrel{+}= [q, e]$\\
\textbf{return} \textsc{fail}\\
\end{tabular}
\end{minipage}%
\hfill
\begin{minipage}[t]{0.34\textwidth}
\centering
\footnotesize
\textbf{Algorithm 2} \approach{2}: Structured intermediate.\label{alg:a2}\par\smallskip
\begin{tabular}{@{}l@{}}
\textbf{Input:} NL task $t$ \quad \textbf{Output:} Result or failure\\[2pt]
$\mathit{msgs} \gets [\text{sys}(\text{schema}), \text{usr}(t)]$\\
\textbf{for} $i \gets 1$ \textbf{to} $3$ \textbf{do}\\
\quad $j \gets \mathrm{parse}(\mathrm{LLM}(\mathit{msgs}))$\\
\quad \textbf{if} invalid \textbf{then} $\mathit{msgs} \mathrel{+}= [j, \mathit{err}]$; \textbf{continue}\\
\quad $q \gets \mathrm{mapper}(j)$\\
\quad \textbf{return} $\mathrm{Joern}(q)$\\
\textbf{return} \textsc{fail}\\
\end{tabular}
\end{minipage}%
\hfill
\begin{minipage}[t]{0.31\textwidth}
\centering
\footnotesize
\textbf{Algorithm 3} \approach{3}: Agentic loop.\label{alg:a3}\par\smallskip
\begin{tabular}{@{}l@{}}
\textbf{Input:} NL task $t$ \quad \textbf{Output:} Answer or failure\\[2pt]
$\mathit{msgs} \gets [\text{sys}(\text{tools}), \text{usr}(t)]$\\
\textbf{for} $s \gets 1$ \textbf{to} $10$ \textbf{do}\\
\quad $r \gets \mathrm{LLM}(\mathit{msgs})$\\
\quad \textbf{if} tool call \textbf{then} exec; append\\
\quad \textbf{else if} answer \textbf{then return} answer\\
\quad \textbf{else return} \textsc{fail}\\
\textbf{return} \textsc{max steps}\\
\end{tabular}
\end{minipage}
\caption{The three approach procedures, side by side. Grey boxes in \Cref{fig:overview} correspond to LLM-mediated steps; white boxes to deterministic steps.}
\label{fig:algorithms}
\end{figure*}

This is the most direct approach, and the one with the largest output space.
The LLM must handle Scala method chaining, the CPG node type hierarchy, and traversal semantics, none of which it has seen much of during training.
To avoid benchmark leakage, all examples in the prompt are drawn from Joern documentation and are distinct from the 20 benchmark tasks.

\subsubsection{Approach 2: Structured Intermediate}\label{sec:a2}

The LLM produces a JSON object that conforms to a predefined schema.
The schema captures the parameters of the analysis task---query type, scope, filters, output columns, flow endpoints---but not the \cpgql{} syntax.
A deterministic mapper covering all benchmark tasks, translates the JSON to the correct query.

For example, given ``trace how user input reaches the database query
in \texttt{processOrder},'' the LLM produces:

\begin{lstlisting}[caption={Example JSON for a data flow query.},label={lst:json-example}]
{
  "query_type": "data_flow",
  "source": {
    "kind": "parameter",
    "method": "processOrder"
  },
  "sink": {
    "kind": "call",
    "name": "execute"
  },
  "output_columns": ["code", "lineNumber"]
}
\end{lstlisting}

The mapper selects the appropriate CPGQL traversal template, fills in
the entity references, and sends the query to Joern. Note that the
natural language request mentions ``database query'' while the JSON
specifies \texttt{execute}---the LLM must infer that database queries
in Java are typically invoked via JDBC methods such as
\texttt{PreparedStatement.execute()}. The LLM's task thus narrows to
selecting among a small set of typed fields and extracting or inferring
the relevant entity names. Some domain knowledge is still required, but
this inference is confined to filling a well-typed field rather than
being entangled with traversal syntax and method chaining. Crucially,
the LLM never needs to relate an identifier to the syntactic structure
of a CPGQL expression, which is precisely the cross-category reasoning
that Anand et al.~\cite{anand2024critical} showed code-LLMs fail to
encode reliably.

\subsubsection{Approach 3: Tool-Augmented Agentic}\label{sec:a3}

Code analysis capabilities are exposed as tools via the HuggingFace function-calling API.
From the model's perspective, this interface is equivalent to MCP~\cite{mcp2024} and similar tool-use protocols: tool schemas are provided as structured descriptions in the system context, and tool results are injected as conversation turns.
The model sees the same sequence of descriptions, invocations, and results regardless of the underlying transport.
The implementations differ in how they parse the model's tool-call output; we discuss this distinction in \Cref{sec:threats}.

\begin{sloppypar}
Five tools cover common analysis operations:
\texttt{find\_methods},
\texttt{find\_calls},
\texttt{trace\_data\_flow},
\texttt{find\_reachable\_by},
and \texttt{run\_custom\_query}.
\end{sloppypar}
The LLM operates in a ReAct-style loop~\cite{yao2023react}: observe request, select tool, observe result, decide next action.

Each step requires the LLM to make a correct tool-selection decision with correct arguments. Unlike \approach{1} and \approach{2}, which each produce a single CPGQL query whose output is compared to the ground truth, \approach{3} runs multiple CPGQL queries internally through its tool calls and synthesises their results into a final text answer.
There is no single generated query to evaluate; the comparison is between this text answer and the Joern output of the ground truth query.
Even with high per-step accuracy, errors compound: a task requiring four tool invocations at 90\% per-step accuracy yields roughly 66\% end-to-end success.
This is consistent with MCP-Universe~\cite{mcpuniverse2025}, where longer task chains show steeper performance drops.
\section{Evaluation}\label{sec:evaluation}

The three approaches are evaluated on a benchmark of code analysis tasks executed against open-source Java projects using Joern.
All experiments use open-weight models accessed through the HuggingFace Inference API.

\subsection{Benchmark}\label{sec:benchmark}

The benchmark consists of 20 tasks organised in three tiers of increasing complexity. Tasks target two real-world Java projects: Apache Commons Lang (a utility library with rich structural patterns) and OWASP WebGoat (a deliberately vulnerable web application with security-relevant data flows).
Nine tasks target Commons Lang; eleven target WebGoat.

Before each experiment run, the Joern server imports the target project's Java source tree and builds a code property graph in memory.
All generated queries---whether produced directly by the LLM (\approach{1}), by the deterministic mapper (\approach{2}), or by the agent's tool calls (\approach{3})---execute against this live CPG.
Result match is determined by running both the generated and the ground truth query against the same graph and comparing their outputs after normalisation (stripping REPL prefixes and collapsing whitespace).
The evaluation is end-to-end: from natural language input to analysis output over real code.

All 20 ground truth queries were manually authored and verified by execution against Joern 4.0.488; an automated validation script confirms that each query runs without error and returns results on the target project.
No benchmark task or its ground truth query appears in any LLM prompt; the examples used in \approach{1} and the schema documentation used in \approach{2} are drawn from Joern's official documentation. The three tiers are as follows:
\begin{enumerate}
    \item \textbf{Structural queries (7 tasks)}: These map to a single \cpgql{} traversal over the AST or type hierarchy.
Examples include listing public methods in a class, finding assignments to a specific variable, and identifying methods with particular annotations.

\item \textbf{Data flow queries (7 tasks)}: These require traversals over the data flow graph using \cpgql{}'s \texttt{reachableBy} and \texttt{reachableByFlows} operators.
Examples include tracing how a method parameter reaches a call site and finding all variables that influence a given condition.

\item \textbf{Composite queries (6 tasks)}:  These combine structural and data flow reasoning, requiring multiple traversals or joint reasoning about control and data flow.
Examples include finding paths from HTTP request parameters to database calls that bypass sanitisation, and identifying methods that transitively call a given sink and accept string parameters.

\end{enumerate}

To our knowledge, no existing benchmark addresses natural-language-to-CPG query translation.
Related benchmarks such as Spider~\cite{spider2018} and BIRD~\cite{bird2023} target SQL over relational databases; CodeSearchNet~\cite{codesearchnet2019} targets code retrieval rather than DSL generation; the OWASP Benchmark evaluates static analysis tools rather than query formulation.
We make this benchmark publicly available as a contribution.\footnote{Repository URL omitted for review.}

\subsection{Models}\label{sec:models}

Each approach is evaluated with four open-weight models arranged in a 2\(\times\)2 design to isolate model-size effects from model-family effects:

\begin{center}
\footnotesize
\begin{tabular}{@{}lll@{}}
  \toprule
  & \textbf{Llama family} & \textbf{Qwen family} \\
  \midrule
  Large (70--72B) & Llama 3.3 70B Instruct & Qwen 2.5 72B Instruct \\
  Small (7--8B)   & Llama 3.1 8B Instruct  & Qwen 2.5 7B Instruct  \\
  \bottomrule
\end{tabular}
\end{center}

All models are accessed through HuggingFace's Inference API with temperature~0 and fixed random seeds (42, 43, 44) for three repetitions, yielding 60 trials per model\(\times\)approach combination.
For \approach{3}, tool schemas are passed via HuggingFace's \texttt{tools} parameter in \texttt{chat\_completion}, which both model families support natively.

\subsection{Metrics}\label{sec:metrics}

Five quantities are measured for each trial:

\paragraph{Result match}
Whether the returned results match the ground truth output, tolerating differences in ordering and whitespace formatting.
This is the most significant metric because syntactically different \cpgql{} queries can produce identical results.

\paragraph{Exact match.}
Whether the generated \cpgql{} string is identical to the ground truth query.
Applicable to \approach{1} and \approach{2} only; \approach{3} produces results directly rather than \cpgql{} strings.

\paragraph{Token consumption.}
Total input and output tokens across all LLM calls for each task, including retries and multi-step interactions.

\paragraph{LLM invocations per task.}
For \approach{2} this is typically~1; for \approach{1} it is 1--4 (with retries); for \approach{3} it varies with the agentic loop length (observed range: 2--10).

\paragraph{Execution success rate.}
Whether the approach produces a result without runtime failure. The definition varies by approach. For \approach{1}, this means the generated CPGQL query executes on Joern without error. For \approach{2}, it additionally requires that the JSON parses, passes schema validation, and maps successfully before Joern execution. For \approach{3}, it means the agentic loop terminates with a final answer rather than exceeding the step limit or failing on a tool-call parse error; individual tool calls execute CPGQL internally, but those successes and failures are not surfaced by this metric. Execution success is therefore a higher bar for \approach{2} than for \approach{1}, and a lower bar for \approach{3}, where it indicates only that the agent produced an answer, not that any particular query ran correctly.

\subsection{Results}\label{sec:results}

A total of 660 trials produced usable data: all 480 trials for \approach{1} and \approach{2} across four models, plus 180 \approach{3} trials for three models.\footnote{\approach{3} for Llama~70B is excluded because 36 of 60 trials failed at the infrastructure level (HuggingFace API rate limits and tool-call parsing errors), leaving too few completed trials for meaningful comparison. The 24 trials that did complete are broadly consistent with the pattern reported here. See \Cref{sec:threats}.}
This section reports findings organised by theme.

\subsubsection{Finding 1: Structured intermediates outperform direct generation}\label{sec:f1}

\Cref{tab:main} presents the main results.
\approach{2} achieves the highest result match rate for every model tested, with the advantage being most pronounced for the large models: +15.0\,pp for Qwen~72B and +25.0\,pp for Llama~70B.
For the small models, the improvement is modest (+3.3\,pp for Qwen~7B, +5.0\,pp for Llama~8B).

\begin{table}[t]
  \centering
  \caption{Result match rate (\%) and execution success rate (\%) across approaches and models. 60 trials per cell (20 tasks \(\times\) 3 reps). Best result match per model in bold. \approach{3} for Llama~70B excluded due to infrastructure failures (see text).}
  \label{tab:main}
  \footnotesize
  \setlength{\tabcolsep}{3.5pt}
  \begin{tabular}{@{}ll rr rr rr@{}}
    \toprule
    & & \multicolumn{2}{c}{\textbf{\approach{1}}} & \multicolumn{2}{c}{\textbf{\approach{2}}} & \multicolumn{2}{c}{\textbf{\approach{3}}} \\
    \cmidrule(lr){3-4} \cmidrule(lr){5-6} \cmidrule(lr){7-8}
    \textbf{Family} & \textbf{Scale} & Res. & Exec. & Res. & Exec. & Res. & Exec. \\
    \midrule
    Qwen  & 72B & 43.3 & 98.3 & \textbf{58.3} & 100  & 25.0 & 90.0 \\
    Llama & 70B & 30.0 & 100  & \textbf{55.0} & 100  & \multicolumn{2}{c}{---\textsuperscript{a}} \\
    Qwen  & 7B  & 31.7 & 100  & \textbf{35.0} & 65.0 & 15.0 & 88.3 \\
    Llama & 8B  & 30.0 & 100  & \textbf{35.0} & 53.3 & 15.0 & 100 \\
    \bottomrule
    \multicolumn{8}{@{}l}{\textsuperscript{a}\footnotesize Excluded; see footnote above.}
  \end{tabular}
\end{table}

\begin{figure}[t]
\centering
\begin{tikzpicture}
\begin{axis}[
    ybar,
    bar width=8pt,
    width=\columnwidth,
    height=5.5cm,
    ylabel={Result match (\%)},
    symbolic x coords={Qwen 72B, Llama 70B, Qwen 7B, Llama 8B},
    xtick=data,
    x tick label style={font=\footnotesize, rotate=15, anchor=east},
    ymin=0, ymax=70,
    legend style={at={(0.98,0.98)}, anchor=north east, font=\footnotesize},
    legend columns=3,
    nodes near coords,
    every node near coord/.append style={font=\tiny},
    enlarge x limits=0.2,
]
\addplot[fill=black!30] coordinates {(Qwen 72B,43.3) (Llama 70B,30.0) (Qwen 7B,31.7) (Llama 8B,30.0)};
\addplot[fill=black!60] coordinates {(Qwen 72B,58.3) (Llama 70B,55.0) (Qwen 7B,35.0) (Llama 8B,35.0)};
\addplot[fill=black!10] coordinates {(Qwen 72B,25.0) (Llama 70B,0) (Qwen 7B,15.0) (Llama 8B,15.0)};
\legend{\approach{1}, \approach{2}, \approach{3}}
\end{axis}
\end{tikzpicture}
\caption{Result match rate by approach and model. \approach{2} (structured intermediate) leads across all models. \approach{3} for Llama~70B is omitted (infrastructure failures).}
\label{fig:main-bar}
\end{figure}
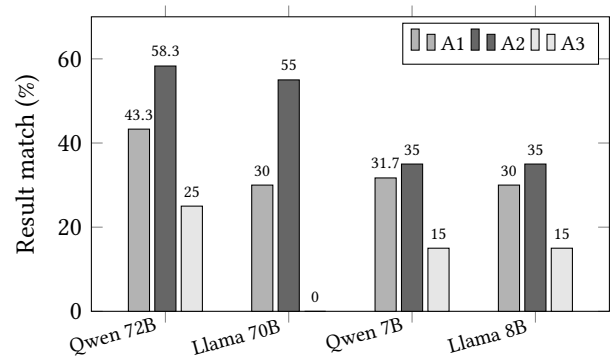

\Cref{fig:main-bar} visualises this pattern.
The structured intermediate consistently outperforms direct generation, and the gap widens with model scale.
\approach{3} underperforms both alternatives on every model.

\begin{findingbox}
Takeaway: The structured intermediate (\approach{2}) outperforms direct generation by 15--25\,pp on large models and 3--5\,pp on small models. The ranking \approach{2}\,>\,\approach{1}\,>\,\approach{3} holds across all models.
\end{findingbox}

\subsubsection{Finding 2: Model size interacts with approach}\label{sec:f2}

The benefit of the structured intermediate depends on model scale.
For large models, \approach{2} provides a substantial improvement because these models are capable enough to fill the JSON schema correctly---both Qwen~72B and Llama~70B achieve 100\% execution success on \approach{2}, meaning every LLM output was valid JSON conforming to the schema that the mapper could translate.

For small models, the picture is different.
Qwen~7B achieves only 65.0\% execution success on \approach{2}, and Llama~8B only 53.3\%.
Nearly half of their outputs fail JSON parsing or schema validation before reaching the mapper.
This creates a ceiling: even though \approach{2}'s architecture is sound, the small models cannot reliably produce the constrained output it requires.

By contrast, \approach{1} achieves near-perfect execution success (98.3--100\%) across all models, because the threshold for ``valid \cpgql{}'' that Joern will attempt to run is lower than the threshold for ``valid JSON conforming to a typed schema.''
The small models write syntactically plausible \cpgql{} that executes but returns wrong results; \approach{2} forces them to be precise about intent, and they often fail at that precision.

Note that \approach{3}'s high execution success rates (88--100\%) are not directly comparable to those of \approach{1} and \approach{2}: they reflect whether the agent produced a final answer, not whether a generated query ran on Joern. Llama~8B achieves 100\% execution success on \approach{3} but only 15.0\% result match, confirming that this metric measures a different property for the agentic approach.

\begin{figure}[t]
\centering
\begin{tikzpicture}
\begin{axis}[
    ybar,
    bar width=6pt,
    width=\columnwidth,
    height=5.5cm,
    ylabel={Rate (\%)},
    symbolic x coords={Q72-A1, Q72-A2, L70-A1, L70-A2, Q7-A1, Q7-A2, L8-A1, L8-A2},
    xtick=data,
    x tick label style={font=\tiny, rotate=30, anchor=east},
    ymin=0, ymax=110,
    legend style={at={(0.02,0.98)}, anchor=north west, font=\footnotesize},
    enlarge x limits=0.1,
]
\addplot[fill=black!20] coordinates {(Q72-A1,98.3) (Q72-A2,100) (L70-A1,100) (L70-A2,100) (Q7-A1,100) (Q7-A2,65.0) (L8-A1,100) (L8-A2,53.3)};
\addplot[fill=black!60] coordinates {(Q72-A1,43.3) (Q72-A2,58.3) (L70-A1,30.0) (L70-A2,55.0) (Q7-A1,31.7) (Q7-A2,35.0) (L8-A1,30.0) (L8-A2,35.0)};
\legend{Execution success, Result match}
\end{axis}
\end{tikzpicture}
\caption{Execution success vs.\ result match for \approach{1} and \approach{2}. Small models (Q7, L8) show high execution success on \approach{1} but low result match; on \approach{2} their execution success drops but result match improves slightly.}
\label{fig:exec-vs-match}
\end{figure}
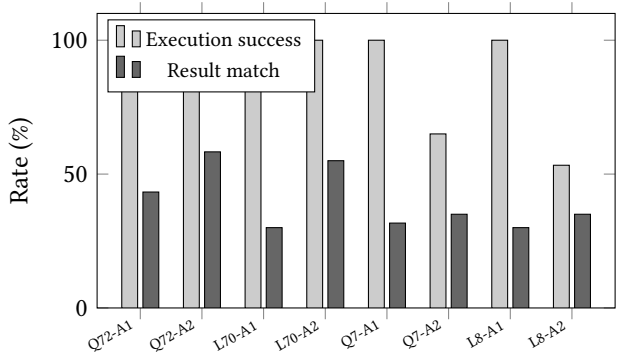

\Cref{fig:exec-vs-match} illustrates this tradeoff.

\begin{findingbox}
Takeaway: Large models fill the JSON schema reliably (100\% execution success) and reap the full architectural benefit. Small models fail schema validation in 35--47\% of attempts, creating a ceiling that limits the structured approach's advantage.
\end{findingbox}

\subsubsection{Finding 3: Agentic generation underperforms at higher cost}\label{sec:f3}

\approach{3} achieves the lowest result match rates across all three models with clean data: 25.0\% for Qwen~72B, 15.0\% for Qwen~7B, and 15.0\% for Llama~8B.
For every model, \approach{3} falls below both \approach{1} and \approach{2}.
Even a relaxed match criterion that tolerates partial overlap raises the numbers only modestly (35.0\%, 15.0\%, and 20.0\% respectively).

The small models complete the agentic loop more quickly---averaging 3.1 steps (Qwen~7B) and 2.8 steps (Llama~8B) compared to 4.8 for Qwen~72B---but this reflects shallower exploration rather than efficiency.
They tend to produce a final answer after fewer tool calls, often without gathering sufficient information.

On Qwen~72B, 8 of 60 trials reached the maximum of 10 steps without producing a final answer.
The agent explored tool results without converging on a conclusion.
The average trial used 4.8 steps and 3.9 tool calls, with an average latency of 32.1 seconds---compared to under 5 seconds for \approach{1} and \approach{2}.

\begin{findingbox}
Takeaway: Agentic generation achieves the lowest accuracy across all models (15--25\%) while taking 6\(\times\) longer per task. Small models exit the loop quickly without gathering sufficient information; large models explore without converging.
\end{findingbox}

\subsubsection{Finding 4: Token cost does not predict accuracy}\label{sec:f4}

\begin{figure}[t]
\centering
\begin{tikzpicture}
\begin{axis}[
    width=\columnwidth,
    height=5.5cm,
    ylabel={Total tokens per task},
    xtick={1,2,3},
    xticklabels={\approach{1}, \approach{2}, \approach{3}},
    ymin=0, ymax=46000,
    ytick={0,5000,10000,15000,20000,25000,30000,35000,40000,45000},
    yticklabel style={font=\tiny},
    scaled y ticks=false,
    enlarge x limits=0.3,
    boxplot/draw direction=y,
]
\addplot+[boxplot prepared={
    lower whisker=1420,
    lower quartile=1431,
    median=1438,
    upper quartile=1449,
    upper whisker=6793},
    fill=black!30, draw=black] coordinates {};
\addplot+[boxplot prepared={
    lower whisker=1583,
    lower quartile=1595,
    median=1608,
    upper quartile=1620,
    upper whisker=3355},
    fill=black!60, draw=black] coordinates {};
\addplot+[boxplot prepared={
    lower whisker=3081,
    lower quartile=4768,
    median=6756,
    upper quartile=21883,
    upper whisker=42790},
    fill=black!10, draw=black] coordinates {};
\end{axis}
\end{tikzpicture}
\caption{Distribution of total token consumption per task (Qwen~72B). \approach{1} and \approach{2} cluster tightly below 2{,}000 tokens. \approach{3} spans a 14\(\times\) range (3{,}081--42{,}790) with a median of 6{,}756---consuming 4\(\times\) more tokens at the median and 8\(\times\) more on average, yet achieving the lowest accuracy.}
\label{fig:tokens}
\end{figure}
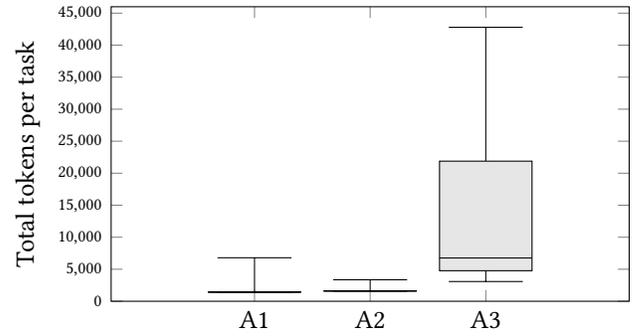

\Cref{fig:tokens} compares the distribution of total token consumption across approaches for Qwen~72B.
The box plot makes the variance story vivid: \approach{1} and \approach{2} are barely distinguishable at the bottom of the scale, each clustered in a narrow band below 2{,}000 tokens, while \approach{3} sprawls across a 14\(\times\) range from 3{,}081 to 42{,}790.
At the median, \approach{3} consumes 4\(\times\) more tokens than \approach{2}; at the mean, 8\(\times\) more.
The interquartile range of \approach{3} alone (4{,}768--21{,}883) is wider than the entire range of \approach{1} and \approach{2} combined.
This variance reflects the unpredictability of the agentic loop: some tasks resolve in two tool calls, others hit the 10-step ceiling without converging.

\begin{findingbox}
Takeaway: \approach{3} consumes 8\(\times\) more tokens on average than \approach{2}, with a 14\(\times\) range in per-task cost, yet achieves 33\,pp lower accuracy. More compute does not compensate for architectural mismatch.
\end{findingbox}

\subsubsection{Finding 5: Tier difficulty varies by approach}\label{sec:f5}

\Cref{tab:tiers} shows the per-tier breakdown for Qwen~72B across all three approaches.
Structural queries are the most accessible tier across all approaches.
The tier difficulty ordering varies across approaches.
For \approach{1}, composite queries are more accessible than pure data flow queries, suggesting that the structural component provides grounding.
For \approach{2}, data flow queries are the most accessible tier (71.4\%), likely because the schema explicitly captures flow source and sink endpoints, reducing the task to classification.
Composite queries are the hardest tier for \approach{2} (44.4\%), which may reflect the difficulty of jointly specifying multiple schema fields.

For \approach{3}, the pattern is different.
The agent handles structural queries reasonably (38.1\%), but its performance on data flow and composite queries drops sharply (19.0\% and 16.7\%).
On the smaller models, \approach{3} solves \emph{only} structural queries: Qwen~7B achieves 42.9\% on structural tasks but 0\% on both data flow and composite tasks.
This suggests that multi-step tool use collapses when the task requires coordinating information across tool calls.

\begin{findingbox}
Takeaway: \approach{3} solves only structural queries on small models (0\% on data flow and composite). The agentic loop fails when tasks require coordinating information across multiple tool calls.
\end{findingbox}

\begin{table}[t]
  \centering
  \caption{Result match by tier (Qwen~72B, 3 reps). Structural: 21 trials (7 tasks \(\times\) 3); Data flow: 21; Composite: 18 (6 \(\times\) 3).}
  \label{tab:tiers}
  \footnotesize
  \begin{tabular}{@{}l ccc ccc ccc@{}}
    \toprule
    & \multicolumn{3}{c}{\textbf{\approach{1}}} & \multicolumn{3}{c}{\textbf{\approach{2}}} & \multicolumn{3}{c}{\textbf{\approach{3}}} \\
    \cmidrule(lr){2-4} \cmidrule(lr){5-7} \cmidrule(lr){8-10}
    \textbf{Tier} & \# & of & \% & \# & of & \% & \# & of & \% \\
    \midrule
  Structural & 11 & 21 & 52.4 & 12 & 21 & 57.1 & 8 & 21 & 38.1 \\
    Data flow  &  7 & 21 & 33.3 & 15 & 21 & 71.4 & 4 & 21 & 19.0 \\
    Composite  &  8 & 18 & 44.4 &  8 & 18 & 44.4 & 3 & 18 & 16.7 \\
    \midrule
    \textbf{All} & \textbf{26} & \textbf{60} & \textbf{43.3} & \textbf{35} & \textbf{60} & \textbf{58.3} & \textbf{15} & \textbf{60} & \textbf{25.0} \\
    \bottomrule
  \end{tabular}
\end{table}

\subsubsection{Finding 6: No complementary coverage from agentic generation}\label{sec:f6}

A natural hope for tool-augmented generation is that it might solve different tasks than the other approaches---perhaps succeeding where structured intermediates fail, making a combined system worthwhile.
The data rule this out.

\Cref{tab:heatmap} shows the per-task breakdown for Qwen~72B across all three approaches.
Each cell reports how many of three repetitions produced a correct result.
Two patterns are immediately visible.
First, the set of tasks solved by \approach{3} (shaded cells in the rightmost column) is a strict subset of those solved by \approach{2}---the agentic approach never solves a task that the structured intermediate cannot.
Second, six tasks (C02, C05, D02, D07, S03, S05) are never solved by any approach; manual inspection reveals these involve uncommon \cpgql{} operators or multi-hop reasoning that no model approximates reliably.

This strict-subset relationship holds across all three models with clean \approach{3} data.
On Qwen~72B, \approach{2} solves 13 of 20 tasks at least once; \approach{3} solves 6 of those 13 and no others.
On Qwen~7B, \approach{2} solves 7 tasks; \approach{3} solves 3 of those 7.
On Llama~8B, \approach{2} solves 7 tasks; \approach{3} solves 3 of those 7.
There is no task for which the agentic approach offers unique coverage.

\begin{findingbox}
Takeaway: The tasks solved by \approach{3} are a strict subset of those solved by \approach{2}, across all models. The agentic approach offers no complementary value that would justify its cost in a combined system.
\end{findingbox}

\begin{table}[t]
  \centering
  \caption{Per-task results for Qwen~72B (3 repetitions per cell). Each cell shows the number of repetitions that produced a correct result. Tasks grouped by tier. \ding{51}\,=\,3/3; \textbullet\,=\,1--2/3; ---\,=\,0/3.}
  \label{tab:heatmap}
  \footnotesize
  \setlength{\tabcolsep}{5pt}
  \begin{tabular}{@{}cl ccc@{}}
    \toprule
    \textbf{Tier} & \textbf{Task} & \textbf{\approach{1}} & \textbf{\approach{2}} & \textbf{\approach{3}} \\
    \midrule
    \multirow{7}{*}{\rotatebox{90}{Structural}}
    & S01 & \ding{51} & \ding{51} & \ding{51} \\
    & S02 & \ding{51} & \ding{51} & \textbullet \\
    & S03 & ---        & ---        & ---        \\
    & S04 & \textbullet & \ding{51} & ---        \\
    & S05 & ---        & ---        & ---        \\
    & S06 & \ding{51} & \ding{51} & \ding{51} \\
    & S07 & ---        & ---        & ---        \\
    \midrule
    \multirow{7}{*}{\rotatebox{90}{Data flow}}
    & D01 & ---        & \ding{51} & ---        \\
    & D02 & ---        & ---        & ---        \\
    & D03 & \ding{51} & \ding{51} & \textbullet \\
    & D04 & ---        & \ding{51} & ---        \\
    & D05 & \textbullet & \ding{51} & ---        \\
    & D06 & \ding{51} & \ding{51} & \ding{51} \\
    & D07 & ---        & ---        & ---        \\
    \midrule
    \multirow{6}{*}{\rotatebox{90}{Composite}}
    & C01 & \ding{51} & \textbullet & ---        \\
    & C02 & ---        & ---        & ---        \\
    & C03 & \ding{51} & \ding{51} & \ding{51} \\
    & C04 & ---        & \textbullet & ---        \\
    & C05 & ---        & ---        & ---        \\
    & C06 & \textbullet & \ding{51} & ---        \\
    \bottomrule
  \end{tabular}
\end{table}

\subsubsection{Finding 7: Exact match understates correctness}\label{sec:f7}

The gap between exact match and result match reveals that LLMs frequently write semantically equivalent but syntactically different queries.
For Qwen~72B on \approach{1}, exact match is 26.7\% while result match is 43.3\%---a 16.6\,pp gap.
The LLM traverses the CPG via a different but valid path and arrives at the same result.

For \approach{2}, this gap is smaller because the deterministic mapper produces exactly one canonical \cpgql{} string per JSON specification.
Any syntactic variation in the output reflects variation in the JSON (e.g., different filter orderings that the mapper normalises), not in the \cpgql{} itself.

This finding has methodological implications: evaluations of DSL generation that rely solely on exact string matching understate actual correctness.
Result-based evaluation, while more expensive (it requires a running execution environment), is more appropriate for measuring practical utility.

\begin{findingbox}
Takeaway: Exact string matching understates correctness by 16.6\,pp on \approach{1}. DSL generation should be evaluated by executing queries and comparing results, not by comparing strings.
\end{findingbox}

\section{Discussion}\label{sec:discussion}

\subsection{Effectiveness of Structured Intermediates}\label{sec:d-structured}

The structured intermediate (\approach{2}) outperforms direct generation (\approach{1}) across all four models.
The data are unambiguous on this point: 15\,pp for Qwen~72B, 25\,pp for Llama~70B, and 3--5\,pp for the smaller models.
What requires interpretation is \emph{why}.

One explanation---and we emphasise that this is our reading of the data, not an experimentally isolated claim---is that \approach{2} transforms the LLM's task from code generation to something closer to classification.
Rather than producing arbitrary \cpgql{} traversal code requiring fluency in a Scala-flavoured DSL with method chaining, graph traversal semantics, and niche operator names, the LLM selects from a small set of well-typed fields: five query types, ten output columns, three flow endpoint types.
The deterministic mapper, verified by 22 unit tests, handles all syntactic and semantic details of \cpgql{} generation.

This framing aligns with the findings of Anand et al.~\cite{anand2024critical}, who showed that code-LLMs fail to encode cross-category relations between syntactic tokens and identifiers.
\approach{2} sidesteps this limitation: the LLM never needs to relate the identifier \texttt{processOrder} to the syntactic structure of a \cpgql{} method traversal; it only places \texttt{processOrder} in the correct JSON field.
Whether this is the full explanation or merely one contributing factor, we cannot determine from our experimental design alone.

What we \emph{can} establish is that the magnitude of the improvement is comparable to the gains reported by MoCQ~\cite{mocq2025} from reducing the CodeQL DSL surface, and by IRIS~\cite{iris2025} from restricting the LLM to classification rather than generation.
All three systems arrive at the same architectural principle from different starting points, which lends credibility to the pattern even if the underlying mechanism remains a hypothesis.

\subsection{Error Compounding in Agentic Generation}\label{sec:d-agentic}

The tool-augmented agentic approach (\approach{3}) performs worst across all models for which we have clean data.
On Qwen~72B, \approach{3} achieves 25.0\% result match compared to 58.3\% for \approach{2}---less than half.
On the smaller models the gap narrows in absolute terms but not in the ordering: Qwen~7B reaches 15.0\% and Llama~8B reaches 15.0\%, both below their \approach{1} scores.

Three failure patterns are consistently observed.
First, the agent selects a tool that returns a large but partially relevant result set, then reasons incorrectly over that set in subsequent steps.
Second, the agent enters exploratory loops, calling tools to gather context rather than to answer the question, consuming steps without converging.
Third, the agent occasionally produces a correct intermediate result from a tool call but misinterprets or reformats it when composing the final answer.

A striking finding is that \approach{3} never solves a task that \approach{2} cannot (\Cref{sec:f6}).
Across all three models with clean data, the set of tasks that \approach{3} answers correctly is a strict subset of those that \approach{2} answers correctly.
The agentic approach does not even offer complementary coverage---it provides no unique capability that could justify its substantially higher cost.

These findings are consistent with MCP-Universe~\cite{mcpuniverse2025}, where task success degrades with chain length, and with QLCoder~\cite{qlcoder2025}, where unrestricted tool access degraded performance.
The compounding-error hypothesis---that even a modest per-step error rate accumulates across multi-step interactions---fits the data, though we note that the hypothesis is structural rather than experimentally isolated.
With an observed average of 4.8 steps per trial on Qwen~72B, even 90\% per-step accuracy would yield roughly 59\% end-to-end success, consistent with the observed gap.

\subsection{Threats to Validity}\label{sec:threats}

The benchmark covers 20 tasks across three complexity tiers.
With 60 trials per model\(\times\)approach cell, the differences between \approach{1} and \approach{2} on large models exceed 15\,pp and are consistent across both model families, but we do not claim comprehensive coverage of all possible analysis tasks.

The benchmark is author-constructed due to the absence of existing NL-to-CPG benchmarks.
To mitigate bias, all ground truths are machine-validated by execution against Joern, and no benchmark content appears in any LLM prompt.
Expanding the benchmark is future work; the current version is publicly released for others to extend.

The deterministic mapper in \approach{2} is hand-written for the schema defined here.
Tasks outside the schema would require schema extension, which is manual work.
The mapper's coverage defines the ceiling of what \approach{2} can express.

\approach{3} uses HuggingFace's function-calling implementation rather than MCP directly.
From the model's input perspective, the two are equivalent: the same tool schemas and result injections appear in the context window.
They differ in output parsing: HuggingFace's parser rejected some of Llama~70B's tool-call formatting attempts (the model generated \texttt{<function=...>} syntax instead of JSON), contributing to infrastructure failures that led us to exclude Llama~70B from the \approach{3} analysis.
An MCP implementation might parse differently, though this would affect only the transport layer, not the model's reasoning.
Comparing tool-use protocol implementations is orthogonal to the question of how much to involve the LLM and is a direction for future work.

Results are specific to the four open-weight models tested.
However, the relative ordering (\approach{2} outperforming \approach{1}, which outperforms \approach{3}) holds across both families at both scales for which we have clean data, which suggests the pattern is not model-specific.

All experiments use temperature~0.
Higher temperatures may produce different variance characteristics but are unlikely to change the central finding about output-space constraint.

The comparison procedure differs slightly for \approach{3}.
For \approach{1} and \approach{2}, both the generated and ground truth queries execute on Joern, and the two outputs are compared directly.
For \approach{3}, the agent produces a text-rendered final answer (e.g., \texttt{List("foo", "bar")}) rather than a \cpgql{} string; this text is compared against the Joern output of the ground truth query after the same normalisation.
The relaxed match metric (which extracts quoted strings and compares as sets) partially addresses formatting differences, but the comparison is inherently noisier than Joern-to-Joern output comparison.

For small models (7--8B parameters), the bottleneck shifts from query correctness to schema compliance.
Guided decoding or constrained generation~\cite{structeval2025} may address this gap by enforcing JSON schema compliance at the token level, effectively removing the schema-compliance burden from the model.
This is a direction for future work.

\section{Related Work}\label{sec:related}

\subsection{LLM-Assisted Static Analysis}\label{sec:rw-sa}

Recent systems that combine LLMs with static analysis tools differ primarily in what the LLM produces and how much of the pipeline it controls.

At one end of the spectrum, IRIS~\cite{iris2025} restricts the LLM to classifying APIs as taint sources or sinks.
These labels are fed into CodeQL via templates; the LLM never writes query code.
On CWE-Bench-Java, IRIS with GPT-4 detects 55 vulnerabilities compared to CodeQL's 27, with fewer false positives.
The deliberate restriction of the LLM's role is a design choice that pays off, though the authors do not frame it as such.

At the other end, QLCoder~\cite{qlcoder2025} asks the LLM to synthesise complete CodeQL queries using an agentic framework with tool access.
It achieves 100\% compilation and 53.4\% success on CWE-Bench-Java.
The authors report that giving the agent unrestricted access to compile-and-run degraded performance because the LLM overused it.
This is a concrete data point in favour of constraining LLM involvement.

QLPro~\cite{qlpro2025} uses a three-role mechanism (Writer, Repair, Execute) to fix syntax errors in generated queries.
CQLLM~\cite{cqllm2025} combines RAG with LoRA fine-tuning.
Both implicitly acknowledge that LLMs produce unreliable DSL code, then add corrective layers rather than questioning whether generation is the right task.

MoCQ~\cite{mocq2025} takes a middle position: it extracts a reduced subset of the CodeQL DSL and provides only that subset as context.
Providing the full specification overwhelmed the model; the subset was tractable.
This is output-space reduction applied at the prompt level.

None of these systems compares architectures with different levels of LLM involvement in a controlled setting.
Each picks a point on the spectrum and optimises it.
While these works target CodeQL specifically, the architectural question generalises.
This paper instantiates it for Joern's \cpgql{}, a DSL with even less representation in LLM training corpora, which makes the output-space argument sharper.

\subsection{What Code-LLMs Fail to Encode}\label{sec:rw-codenot}

The hypothesis that LLMs are unreliable generators of static analysis queries has empirical grounding beyond anecdotal compilation failures.
Anand et al.~\cite{anand2024critical} analysed attention maps and hidden representations of code-LLMs at the token level.
They categorised tokens into syntactic tokens (keywords, operators, delimiters) and identifiers (variable names, function names), and found that models encode relations within each category but fail to encode relations across them.
Paradoxically, larger models encoded less structural information than smaller ones.
Fine-tuned models performed worse than their pre-trained counterparts on these structural probes.

These findings matter directly for DSL generation.
A typical \cpgql{} query---e.g., one that retrieves the parameters of a method named \texttt{foo}---requires the model to connect the identifier \texttt{foo} to the syntactic structure of a method traversal.
If code-LLMs cannot reliably represent these cross-category links, generating correct DSL queries from natural language is inherently fragile, regardless of model scale.

\subsection{Structured Output from LLMs}\label{sec:rw-structured}

LLMs struggle with the syntax of domain-specific languages, and this is not specific to any single tool.
StructEval~\cite{structeval2025} evaluates LLMs across 18 structured formats and reports gaps even for frontier models.
Work on JSON processing shows that generating code to parse JSON outperforms direct inspection by 3--50\%, suggesting LLMs handle structured data better when the task is constrained.

The implication is direct: reducing the LLM's task from ``generate \cpgql{}'' to ``produce a JSON object conforming to a known schema'' moves the problem to a representation where LLMs are more competent.
Schema-constrained JSON output is now a standard feature of major LLM APIs and can be enforced through guided decoding on open-weight models, further reducing failure modes.

The analogy to Text-to-SQL~\cite{texttosql2024} is useful.
That community has studied how schema information and intermediate representations affect query generation.
The intermediate JSON approach presented here is the static-analysis equivalent of schema-linking: generate a constrained intermediate, let deterministic code produce the query.

\subsection{Tool-Augmented LLM Generation}\label{sec:rw-uncertainty}

The ReAct framework~\cite{yao2023react} established the pattern that \approach{3} implements: the LLM alternates between reasoning about a task and invoking external tools, observing results at each step.
Toolformer~\cite{schick2023toolformer} and Gorilla~\cite{patil2023gorilla} explore the same pattern from different angles.
In all cases, what matters from the model's perspective is the tool schema it sees and the result it gets back, not the transport mechanism---whether tools are invoked via MCP, OpenAI function calling, or HuggingFace's \texttt{tools} parameter, the model's context window contains the same sequence of tool descriptions, invocations, and results.

Recent work on uncertainty in tool-augmented LLM systems provides theoretical context for the hypothesis.
Tools in the Loop~\cite{toolsinloop2025} proposes a framework for jointly modelling the predictive uncertainty of the LLM and the external tools, showing that both contribute to overall reliability.
MCP-Universe~\cite{mcpuniverse2025} reports that GPT-5 achieves only 43.72\% overall success in realistic MCP environments, with failures concentrated in content generation rather than format compliance.
CA-MCP~\cite{camcp2025} proposes reducing LLM involvement in MCP workflows by letting servers coordinate without continuous LLM orchestration.
These findings support the general principle: less LLM mediation, better outcomes in structured domains.

\subsection{Code Property Graphs and Joern}\label{sec:rw-cpg}

Code Property Graphs were introduced by Yamaguchi et al.~\cite{joern2014} as a unified representation combining ASTs, CFGs, and DFGs.
Joern implements this representation and provides \cpgql{}, a Scala-based query language for traversing the graph.
The CPG representation has since been adopted by other tools and standardised through efforts such as the Open Source Security Foundation's CPG specification.
Joern supports multiple languages (Java, C/C++, Python, JavaScript) and is actively maintained.

\cpgql{} queries use method chaining over graph traversals.
A typical query reads the CPG, filters nodes by type or name, traverses edges, and collects results.
The language is expressive but niche: it is underrepresented in LLM training data compared to mainstream languages, which makes it a suitable test case for studying LLM limitations in DSL generation.

\section{Conclusion}\label{sec:conclusion}

This paper investigates how much of a static analysis pipeline should be delegated to an LLM.
Three architectures are instantiated along a spectrum of LLM involvement---direct \cpgql{} generation, structured JSON intermediates, and tool-augmented agentic generation---and evaluated on 20 code analysis tasks using four open-weight models.

The structured intermediate representation achieves the highest result correctness across all models, with improvements of 15--25\,pp over direct generation on large models.
The agentic approach performs worst despite consuming 8\(\times\) more tokens: 25.0\% result match on Qwen~72B compared to 58.3\% for the structured approach, and 15.0\% on both smaller models.
The set of tasks solved by the agentic approach is a strict subset of those solved by the structured approach, offering no complementary coverage.
Model size interacts with approach: the benefit of structured intermediates is strongest for large models that can reliably fill the schema, while small models face a bottleneck in schema compliance that limits the architectural advantage.

The principle is not specific to \cpgql{}.
Any domain with a formal schema, typed query language, and deterministic execution can benefit from the same decomposition: let the LLM handle what it handles well---natural language understanding and classification into a schema---and let deterministic code handle the rest.
The key precondition is that the space of valid queries can be captured by a schema compact enough for the LLM to fill reliably, which is not guaranteed at all model scales.

The benchmark and all experimental infrastructure are publicly available to support replication and extension.\footnote{Repository URL omitted for review.}

\section*{Acknowledgements}

The artifact and the data to validate the work's claims are listed are made available and the authors request to be included for the Artifact evaluation badge. 

\paragraph{Artifact.}
The experimental infrastructure, benchmark, raw results, and analysis scripts
accompanying this paper are archived at
\url{https://zenodo.org/records/18888136}.
The artifact includes all 660 trial outputs, the deterministic mapper with its
test suite, and scripts to reproduce the paper's tables and figures from the
included data. We apply for the \emph{Artifacts Evaluated~-- Functional} badge.

\bibliography{references}

\end{document}